\documentclass[useAMS,usenatbib]{mn2e}

\usepackage{graphicx}
\usepackage{amsmath}
\usepackage{txfonts}
\usepackage{natbib}
\usepackage{textcomp}
\usepackage{subfigure}

\title[]{Collisional excitation of singly deuterated ammonia NH$_2$D by H$_2$.}
\author[F. Daniel et al.]{F. Daniel$^{1}$\thanks{E-mail:
fabien.daniel@obs.ujf-grenoble.fr}, 
A. Faure$^{1}$, 
L. Wiesenfeld$^{1}$,
E. Roueff$^{2}$,
D.C. Lis$^{3,4}$,
P. Hily--Blant$^{1}$
\footnotemark[1] \\
$^{1}$ IPAG, Observatoire de Grenoble, Universit\'e Joseph Fourier, CNRS UMR5571, B.P. 53, 38041 Grenoble Cedex 09, France \\
$^{2}$ LERMA and UMR 8112, Observatoire de Paris, Place J. Janssen, 92190 Meudon, France \\
$^{3}$ Sorbonne Universit\'{e}s, Universit\'{e} Pierre et Marie Curie, Paris 6, CNRS, Observatoire de Paris, UMR 8112, LERMA,  Paris, France \\
$^{4}$ California Institute of Technology, Cahill Center for Astronomy and Astrophysics 301-17, Pasadena, CA 91125, USA}
\begin{document}

\date{Accepted XXX. Received XXX; in original form XXX}

\pagerange{\pageref{firstpage}--\pageref{lastpage}} \pubyear{2014}

\maketitle

\label{firstpage}

\begin{abstract}
The availability of collisional rate coefficients with H$_2$ is a pre-requisite for interpretation of
observations of molecules whose energy levels are populated under
non local thermodynamical equilibrium conditions. In the current study, we present collisional 
rate coefficients for the NH$_2$D / para--H$_2$($J_2 = 0,2$) collisional system, for energy levels up to 
$J_\tau = 7_7$ ($E_u$$\sim$735 K) and for gas temperatures in the range $T = 5-300$K.
The cross sections are obtained using the essentially
exact close--coupling (CC) formalism at low energy and at the highest energies, we used the 
coupled--states (CS) approximation. For the energy levels up to $J_\tau = 4_2$ 
($E_u$$\sim$215 K), the cross sections obtained
through the CS formalism are scaled according to a few CC reference points. 
These reference points are subsequently used to estimate the accuracy of the rate coefficients for higher levels, 
which is mainly limited by the use of the CS formalism. Considering the current potential energy surface, 
the rate coefficients are thus expected to be accurate to within 5\% for the levels below $J_\tau = 4_2$,
while we estimate an accuracy of 30\% for higher levels.
\end{abstract}


\section{Introduction}

Singly deuterated ammonia, NH$_2$D, was first tentatively detected toward the
Kleinmann-Low nebula in Orion \citep{rodriguez1978} and toward Sgr
B2 \citep{turner1978}. However, due to the molecular richness of these objects, a blending
with transitions from other molecular species could not be discarded at that time.
The first unambiguous detection of NH$_2$D was subsequently performed by \citet{olberg1985} in the cold 
environment of L183, S140 and DR21(OH) at 85 GHz and 110 GHz, thanks to the high spectral resolution
of the observations which enabled to resolve the hyperfine structure of the lines.
Since then, its centimeter and millimeter
rotational transitions have been used as tracers of the physical
conditions and chemistry of the molecular gas over a wide range of
conditions, ranging from cold prestellar cores \citep[e.g.][]{tine2000,saito2000,shah2001,hatchell2003, roueff2005, busquet2010} 
to warm star--forming regions \citep[e.g.][]{walmsley1987,pillai2007}.
In cold environments, the deuterium fractionation was derived to be several 10$^{-2}$.
Such high fractionation ratios are within the values predicted by gas phase models at low temperatures \citep{roueff2005}.
In warm environments such as Orion KL, the fractionation ratio was derived to be 0.003 by \citet{walmsley1987} and interpreted as a possible signature of mantle evaporation. More recently, NH$_2$D has been detected at much higher frequencies thanks to Herschel. In Orion KL \citep{neill2013},
several excited levels of NH$_2$D up to $J_{K_a,K_c}$ = 7$_{5,3}$ ($E_u$ $\sim$ 600 K)\footnote{NH$_2$D is an asymmetric top whose
rotational energy structure can be described with the $J$, $K_a$, $K_c$ quantum numbers. Alternatively, 
one may use the pseudo quantum number $\tau = K_a - K_c$. In what follows, we mainly use the $J_\tau$ notation to describe
the rotational energy levels but can alternatively make use of the $J_{K_a,K_c}$ notation, which is usually employed in astrophysical studies.} 
were detected and the fractionnation ratio was derived to be 0.0068.
The temperature and density associated to the emitting regions were estimated to T $\sim$ 100--300K and n(H$_2$) $\sim$ 10$^7$ -10$^8$ cm$^{-3}$.

In order to interpret the NH$_2$D observations, a key ingredient of the modelling are
the collisional rate coefficients. Before the current calculations, the only 
available rate coefficients considered He as a collisional partner \citep{machin2006}.
However, it was shown in the case of ND$_2$H that the rate coefficients with He 
and H$_2$ can differ by factors 3--30, depending on the transition
\citep{machin2007,wiesenfeld2011}. Thus, a dedicated calculation for NH$_2$D with H$_2$
is required in order to interpret the observations of this molecule. Indeed, 
as discussed by \citet{daniel2013}, a simple scaling of the NH$_2$D /  He rate coefficients is 
insufficient to accurately model the observations, at least under cold dark clouds conditions.

The paper is organized as follow. The potential energy surface is described in 
Section \ref{potentiel} and the collisional dynamics based on this surface in Section \ref{dynamique}.
We then describe the rate coefficients in Section \ref{rates} with some emphasis placed on 
their expected accuracy. In Section \ref{discussion}, 
we discuss the current results with respect to other related collisional systems and finally, we 
present conclusions in Section \ref{conclusion}. 

\section[]{Potential Energy Surface (PES)} \label{potentiel}

The rigid-rotor NH$_2$D-H$_2$ potential energy surface (PES) was
derived from the NH$_3$-H$_2$ PES computed by \cite{maret2009}, but in
the principal inertia axes of NH$_2$D. The scattering equations to be
solved (see below) are indeed written for a PES described in the frame
of the target molecule, here NH$_2$D. In the original NH$_3$-H$_2$
PES, the ammonia and hydrogen molecules were both assumed to be rigid,
which is justified at temperatures lower than $\sim$1000~K. The NH$_3$
and H$_2$ geometries were taken at their ground-state average values,
as recommended by \cite{faure2005} in the case of the H$_2$O-H$_2$
system. The rigid-rotor NH$_3$-H$_2$ PES was computed at the
coupled-cluster CCSD(T) level with a basis set extrapolation
procedure, as described in \cite{maret2009} where full details can be
found. In the present work, the geometry of NH$_2$D was assumed to be
identical to that of NH$_3$, i.e. the effect of deuterium substitution
on the ammonia geometry was neglected. This assumption was previously
adopted for the similar ND$_2$H-H$_2$ system by
\cite{wiesenfeld2011}. We note that internal geometry effects are indeed
expected to be only moderate ($\lesssim$ 30\%) at the temperatures
investigated here, as shown by \cite{scribano2010} for the D$_2$O-H$_2$
system. The main impact of the isotopic substitution on the
NH$_3$-H$_2$ PES is therefore the rotation of the principal inertia
axes.

We have thus expressed the NH$_3$-H$_2$ PES of \cite{maret2009} in the
principal inertia axes of NH$_2$D. The transformation can be found in
\cite{wiesenfeld2011} (Eqs.~2-3) where the $\gamma$ angle is equal to
-8.57~degrees for NH$_2$D, as determined experimentally by
\citep{cohen1982}. The shift of the center of mass was also taken into
account: the coordinates of the NH$_2$D center of mass in the
NH$_3$-H$_2$ reference frame were found to be $X_{\rm
  CM}$=0.101312~Bohr and $Y_{\rm CM}$=-0.032810~Bohr, where $X_{\rm
  CM}$ and $Y_{\rm CM}$ were calculated for the ground-state average
geometry of NH$_3$ used by \cite{maret2009}. The NH$_2$D-H$_2$ PES was
generated on a grid of 87~000 points consisting of 3000 random angular
configurations combined with 29 intermolecular distances in the range
3-15~Bohr. This PES was finally expanded in products of spherical
harmonics and rotation matrices, as in \cite{wiesenfeld2011} (see their
Eq.~4), using a linear least-squares fit procedure. We selected
iteratively all statistical significant terms using the procedure of
\cite{rist2011} applied at all intermolecular distances. The final
expansion included anisotropies up to $l_1$=10 for NH$_2$D and $l_2$=4
for H$_2$, resulting in a total of 210 angular basis functions. The
root mean square residual was found to be lower than 1~cm$^{-1}$ for
intermonomer separations $R$ larger than 5~Bohr, with a corresponding
mean error on the expansion coefficients smaller than 1 cm$^{-1}$.

\section[]{Collisional dynamics} \label{dynamique}

In order to describe the NH$_2$D energy structure, we adopted the same approach than
\citet{machin2006}, i.e. we assumed that NH$_2$D is a rigid rotor and we neglected 
the inversion motion corresponding to the tunneling of the nitrogen atom through the H$_2$D
plane.
The spectroscopic constants, adapted from the CDMS catalogue \citep{muller2001} and 
from \citet{coudert1986}, are given in Table 2 of \citet{machin2006}.
As outlined in \citet{machin2006}, neglecting the inversion motion leads to neglecting the 
difference between the ortho and para states of NH$_2$D, which are thus treated as degenerate. 
Indeed, by considering the inversion motion, every rotational state would be split in
two states which are either symmetric or anti--symmetric under exchange
of the protons. By combining the symmetry of each state with the symmetry of the nuclear
wave functions, each state will either correspond to an ortho or para state to ensure that
the overall wavefunction is anti--symmetric under the exchange operation.
Since the collisional transitions are only possible within a given symmetry, the treatment
of the collisional dynamics would be essentially similar for each species, except for the slight 
differences of energy between the ortho and para states,
which is of the order of $\sim$0.4 cm$^{-1}$ independently of the rotational state
\citep{coudert2006}.
Hence, by neglecting the inversion motion, we obtain
a set of rate coefficients which applies to both the ortho and para symmetries of NH$_2$D. 
Finally, we also considered H$_2$ as a rigid rotor and adopted a rotational 
constant of 59.2817 cm$^{-1}$ \citep{huber1979}.
The reduced mass of the collisional system is $\mu$ = 1.812998990 amu.

\begin{figure}
\begin{center}
\includegraphics[angle=0,scale=0.45]{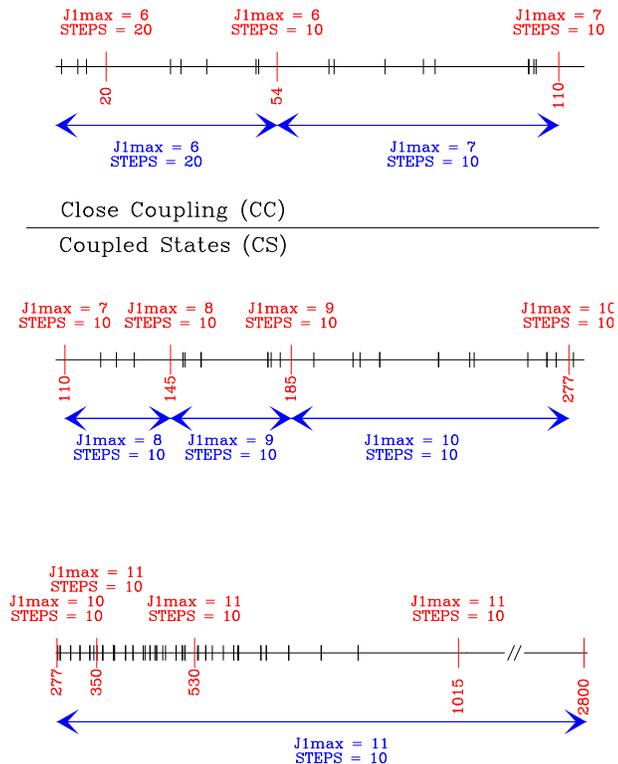}
\caption{Parameters that describe the convergence of the dynamical calculations performed 
with MOLSCAT. J1max stands for the maximum quantum number of NH$_2$D and STEPS is 
inversely proportional to the integration steps of the propagator. In red, we indicate 
the total energies at which we checked for the convergence of these two parameters. 
In blue, we indicate the values used for each energy range. Additionally, we also reported the energy 
levels of NH$_2$D which appear as vertical lines.}
\label{fig:param-conv}
\end{center}
\end{figure}

\begin{figure}
\begin{center}
\includegraphics[angle=270,scale=0.35]{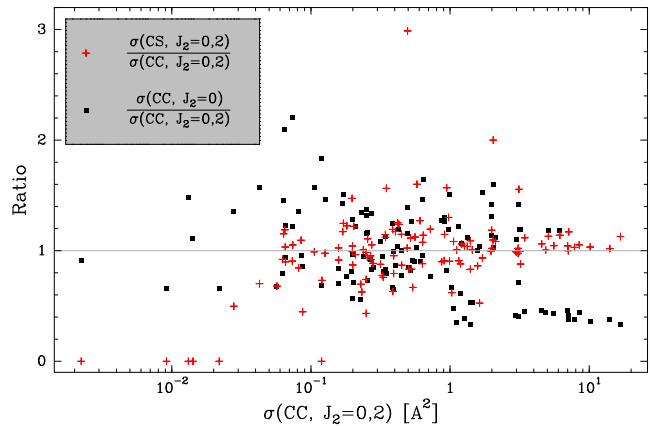}
\caption{Ratio at 110 cm$^{-1}$ between two approximate sets of cross sections with respect to 
CC cross sections obtained with a $J_2=0,2$ basis of H$_2$. The first approximate set (red crosses) 
corresponds to cross sections obtained with the CS formalism and with a $J_2=0,2$ basis of H$_2$. The second
set (black points) is obtained with the CC formalism but the basis of H$_2$ is reduced to $J_2 =0$.}
\label{fig:method}
\end{center}
\end{figure}

In order to solve the collisional dynamics, we used the MOLSCAT 
code\footnote{J. M. Hutson and S. Green, MOLSCAT computer code, version 14 (1994), distributed by 
Collaborative Computational Project No. 6 of the Engineering and Physical Sciences Research Council (UK).}.
Since NH$_2$D is observed in warm media and since the transitions detected include levels up to $J=7$, 
our goal was set accordingly. We performed the calculations in order to provide rate coefficients for the 
NH$_2$D energy levels up to $J_\tau = 7_7$  and for the temperature range $T = 5-300$K.
A first step consisted of adjusting the convergence of the dynamical calculations for these levels and for total 
energies up to $E_t$$\sim$3000 cm$^{-1}$. We thus performed a few calculations
at some specific values of the total energy for which we determined the number of NH$_2$D rotational energy 
levels, as well as the integration step required to insure a convergence better than 1\% below 
530 cm$^{-1}$ and better than 5\% above this threshold. 
These test calculations are summarized in Fig. \ref{fig:param-conv} and the parameters used for each energy range 
are also indicated. 
We found that including the $J_2 = 2$ level of para--H$_2$ in the dynamical calculations had a large impact on 
the resulting cross sections. Indeed, at some specific energies, we found that the cross sections obtained with the 
$J_2 = 0$ or $J_2 = 0,2$ basis can differ by up to a factor 7. An example of the influence of 
the $J_2 = 2$ energy level of H$_2$ on the cross sections is given in Fig. \ref{fig:method}. 
This figure will be further discussed below.
   
\begin{table}
\caption{Step between the consecutive total energies used to characterize the cross sections.}
\begin{center}
\begin{tabular}{cc}
\hline
 Energy range (cm$^{-1}$) & step in energy (cm$^{-1}$) \\ \hline
         $<$ 110 & 0.1 \\
  110 $-$ 277 & 0.2 \\
  277 $-$ 630 & 0.5 \\
  630 $-$ 690 & 2 \\
  690 $-$ 755 & 4 \\
  755 $-$ 1355 & 20 \\
  1355 $-$ 2800 & 50 \\ \hline
\end{tabular}
\end{center}
\label{table:E_step}
\end{table}%

\begin{figure*}
\begin{center}
\includegraphics[angle=270,scale=0.7]{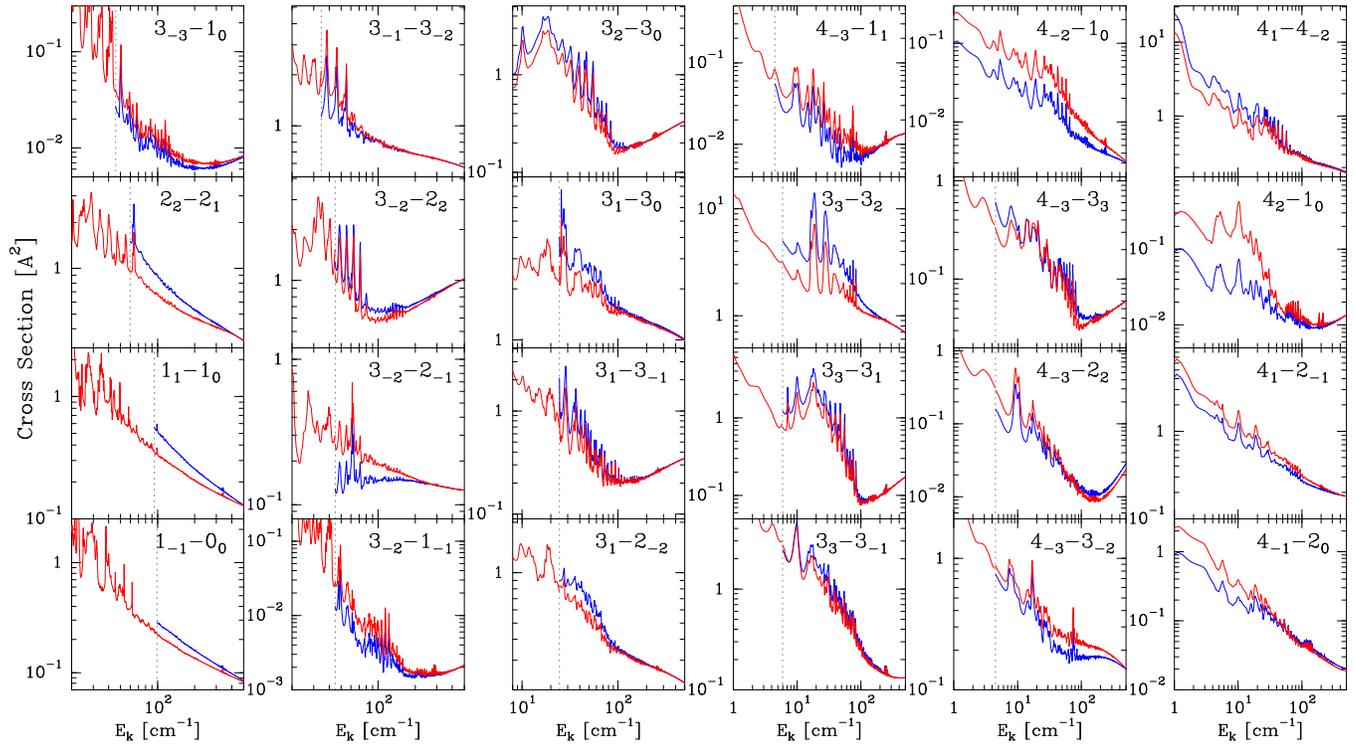}
\caption{Cross sections as a function of the kinetic energy $E_k$. The blue curve corresponds to the CS calculations.
The red curve corresponds to the CC calculations, below a total energy of 110 cm$^{-1}$ and to the scaled CS calculations
above this threshold. The correspondence between this particular value of the total energy and the kinetic energy is 
indicated for each transitions by a dotted vertical line.}
\label{fig:xs_scaled}
\end{center}
\end{figure*}

At low total energy (i.e. $E_t$ $<$ 110 cm$^{-1}$), we performed dynamical calculations
with the accurate close--coupling (CC) formalism. However, because of the increase of 
the computational time with the number of channels, the cost in term of CPU time 
became prohibitive at high energy. As an example, in the range 100--110 cm$^{-1}$, 
the CPU time typically ranged from 40 to 90 hours per energy, the amount of time being dependent 
on how far the energy is from the last opened energy level.
In order to keep the CPU time spent within a reasonable amount of time, we had two options. We could either 
increase the step between two consecutive energy grid points, or switch to an approximate method in order to solve
the collisional dynamics. While the first option could be a good choice for the low 
lying energy levels, it would turn to be a bad choice
for the highest ones, since our calculations show that typically, the cross sections have resonances for kinetic energies up to
$\sim$150 cm$^{-1}$ above the threshold of the transition. Thus, because the NH$_2$D energy levels are close in energy,
and because we are interested in the levels up to J$_\tau$ = $7_7$  (E$\sim$512 cm$^{-1}$), an accurate description of the 
resonances for all the transitions require the step between consecutive energies to be low 
(i.e. typically of 0.2 cm$^{-1}$) up to a total energy of $\sim$660 cm$^{-1}$. 
Therefore, we made the choice to use an approximate method to solve 
the dynamics. We considered two options: reducing the basis of H$_2$ still using 
the CC formalism or using the coupled--states (CS) formalism \citep{mcguire1974}
with the $J_2$ = 0,2 basis for H$_2$. In Fig. \ref{fig:method}, we show the ratio of the cross--sections obtained
with those two treatments with respect to the cross sections obtained with the CC formalism and with the 
$J_2$ = 0,2 basis for H$_2$. In this example, the total energy is 110 cm$^{-1}$ but we checked that the conclusions are 
similar at other values. Considering this figure, it appears that the CS method gives a fairly accurate
description of the dynamics, especially for the transitions with the highest cross sections. Indeed, it can be seen
that for the transitions with cross sections higher than 2 $\AA^2$, the CS method is accurate within a factor $\sim$ 1.2 for most 
transitions. On the other hand, reducing the basis of the H$_2$ molecule with the CC formalism introduces larger errors,
up to a factor $\sim$3 in this case.
We thus made the choice to use the CS approximation to perform the 
dynamical calculations above 110 cm$^{-1}$, 
since the reduced amount of computational time (typically reduced by a factor 10--20) 
enables to adopt a fine energy grid. 
The step between consecutive energies is given in table \ref{table:E_step}. 
In order to emphasize on the necessity to resort to an approximate method to perform
the dynamical calculations, we note that the total CPU time spent to obtain the rate coefficients 
was $\sim$170 000 hours ($\sim$19 years), a time however substantially reduced by the availability 
of a cluster of cores (336 cores of 2.26 GHz). Finally, in order to obtain a set of rate coefficients 
as accurate as possible, 
we scaled the CS cross sections by performing a few calculations with the CC formalism, up to total energies of 
250 cm$^{-1}$. These calculations enable to scale the transitions that involve the levels up to $J_\tau = 4_2$
and the procedure used to perform the scaling is described in the appendix. The scaled and original 
CS cross sections are reported in Fig. \ref{fig:xs_scaled} for a few transitions. Note that as expected from 
formal criteria \citep[see][and references therein]{heil1978},
we observe that the accuracy of the CS formalism increases with the kinetic energy.

\begin{figure*}
\begin{center}
\includegraphics[angle=270,scale=0.7]{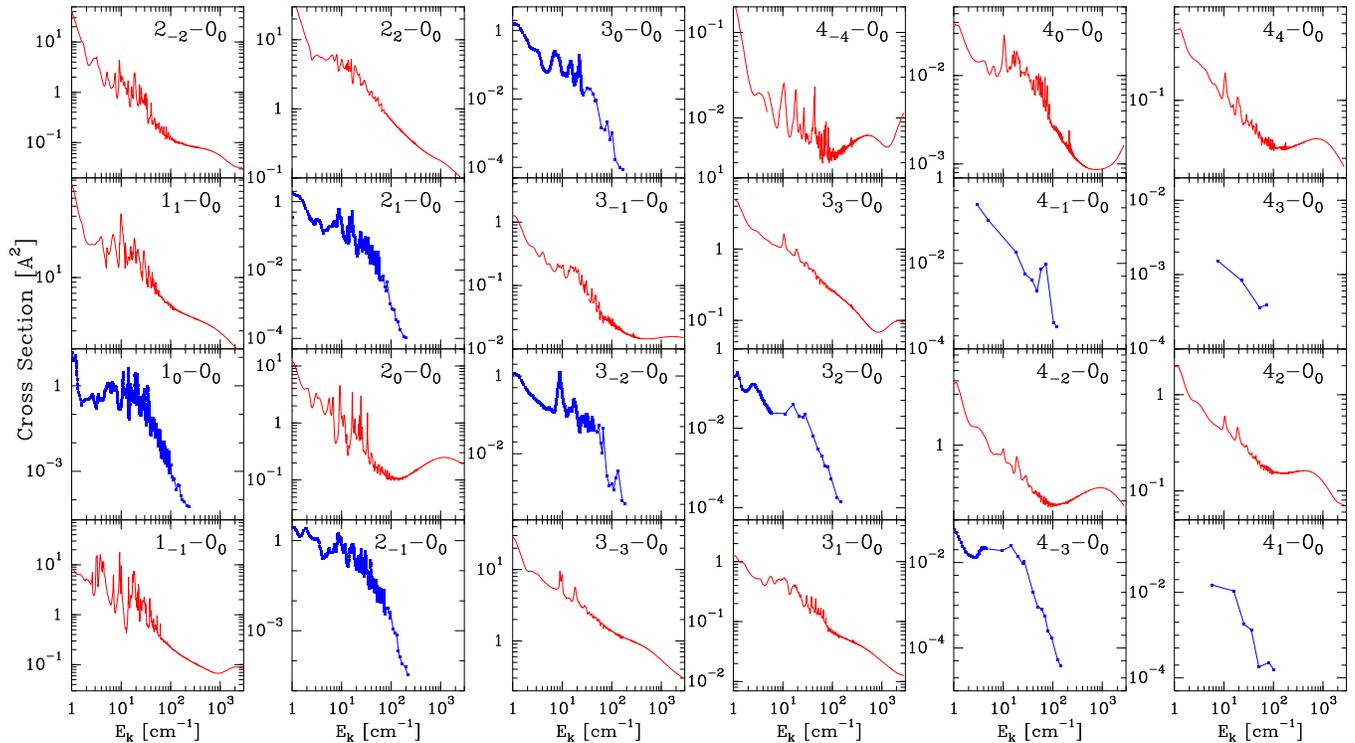}
\caption{Cross sections for the transitions connected to the fundamental energy level and for levels up to 
$J_\tau = 4_4$. The red curves correspond to CC plus scaled CS cross sections. The blue curves correspond to CC 
calculations and are truncated in energy since the CS formalism predict null cross sections.}
\label{fig:CS_null}
\end{center}
\end{figure*}

Apart from its accuracy, the CS formalism has the additional drawback that some transitions are 
predicted to be null while they should not be, which is a consequence of the neglect of small coupling terms.
This can be seen in Fig. \ref{fig:method} where a few transitions 
below 0.2 $\AA^2$ have ratios equal to zero. The transitions which are predicted to be null with the CS
formalism correspond to $J_\tau \to 0_0$ where $\tau$ ranges from $\tau = -J+1$ to 
$\tau = J-1$ in a step of 2. The transitions concerned correspond to H$_2$ 
transitions with $J_2 = 0 \to 0$. In Fig. \ref{fig:CS_null}, we report the de-excitation transitions connected to 
the $0_0$ level for the levels up to $J_\tau = 4_4$. For the transitions with non--null CS 
cross sections (indicated in red in Fig. \ref{fig:method}), the values reported correspond to the 
CC and scaled--CS cross sections. For the other transitions (indicated in blue),  we report the CC points 
we calculated. For the latter, we included the reference CC points used to scale the CS cross sections,
i.e. for total energies between 110 and 250 cm$^{-1}$, which explains why the step in energy is large for the highest 
kinetic energies. First, when considering this figure, it is important to notice that the cross sections predicted to be null
with the CS formalism are indeed of lower magnitude by comparison with other transitions. 
Additionally, from the few transitions
available, it appears that these cross sections decrease quickly with increasing kinetic energy. As an example, the 
$1_0 \to 0_0$ transition decreases by 4 orders of magnitude between 1 cm$^{-1}$ and 110 cm$^{-1}$.
Finally, all these transitions seem to share a similar shape. These peculiarities of the null CS cross sections makes 
it possible to calculate the rate coefficients for the levels up to $J_\tau = 3_{2}$, with a reasonable accuracy up to
300 K, despite the truncation in energy of the grid. 
This point will be further discussed in the next Section where we also derive analytical formulae which are used
to estimate the null $J_\tau \to 0_0$ rate coefficients for the levels above $J_\tau = 3_{2}$.

\begin{figure}
\begin{center}
\includegraphics[angle=270,scale=0.45]{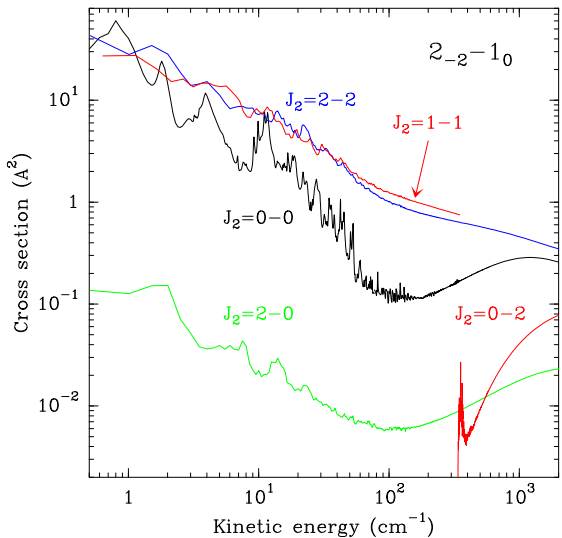}
\caption{Cross sections associated to the $2_{-2} \to 1_0$ NH$_2$D transition and
corresponding to the various $J_2 \to J_2'$ H$_2$ transitions with para--H$_2$
either in its fundamental state ($J_2 = 0$) or first excited state ($J_2 = 2$).
The cross section with ortho--H$_2$ ($J_2 = 1-1$) is also indicated, for a reduced range of kinetic energies.}
\label{fig:XS_J2=2}
\end{center}
\end{figure}

Finally, since we performed calculations with a $J_2 = 0,2$ basis for H$_2$ and since 
the energy grid was well sampled, the cross sections that involve the excited $J_2 = 2$ state of H$_2$
can be used to derive the corresponding rate coefficients. 
Indeed, the convergence criteria previously discussed also apply to the transitions
$J_2 \to J_2'$ with either $J_2$ or $J_2'$ equal to 2. In Fig. \ref{fig:XS_J2=2}, 
we give a characteristic example of such transitions and show the 
transitions related to the $2_{-2} \to 1_0$ state--to--state transition of NH$_2$D and that 
involve the various possible rotational states of H$_2$. From this figure, it appears that the 
transitions which are inelastic for H$_2$ are of lower magnitude than the elastic 
transition, by typically one or two orders of magnitude. Moreover, we find that the 
$J_2 = 2 \to 2$ transitions are larger than the transitions $J_2 = 0 \to 0$. These conclusions
apply to the whole set of transitions and were already reached for other molecular systems 
\citep[see e.g.][]{dubernet2009,daniel2010,daniel2011,wiesenfeld2013}.

\section{Rate coefficients} \label{rates}

The collisional de-excitation rate coefficients are calculated by averaging 
the cross sections with a Maxwell--Boltzmann distribution that describes the 
distribution of velocity of the molecules in the gas \citep[see e.g. eq. (5) in][]{wiesenfeld2011}.
In Table \ref{table:rates_j2=0-0}, we give the de--excitation rate coefficients for levels up to 
$J_{\tau} = 2_2$, with $J_2 = 0 \to 0$, and for temperatures in the range T = 5--100K.
The way these rates are obtained is further discussed below.
This table is directly comparable to Table 7 of \citet{machin2006} where the same 
quantities are given for the NH$_2$D / He system.
The whole set of rate coefficients, with higher temperatures and with a more extended set of 
molecular levels will be made available through the LAMDA \citep{schoier2005} and BASECOL \citep{dubernet2013} databases.

\begin{table*}
\caption{De-excitation rate coefficients for NH$_2$D / H$_2$ (J$_2$ = 0-0) as a function of temperature and for NH$_2$D levels 
up to $J_{\tau} = 2_2$.}
\begin{center}
\begin{tabular}{ccccccc}
\hline \hline
Transition & \multicolumn{6}{c}{Rate coefficients (cm$^3$ s$^{-1}$)} \\
 & 5K & 10K & 25K & 50K & 75K & 100K \\ \hline

$1_{ 0,  1 }  \to 0_{ 0,  0 } $  &         1.24(-11) &         1.17(-11) &         9.58(-12) &         6.86(-12) &         5.41(-12) &         4.58(-12) \vspace{0.2cm} \\
$1_{ 1,  1 }  \to 0_{ 0,  0 } $  &         2.00(-12) &         2.06(-12) &         1.62(-12) &         9.69(-13) &         6.45(-13) &         4.66(-13) \\
$1_{ 1,  1 }  \to 1_{ 0,  1 } $  &         1.17(-10) &         1.20(-10) &         1.16(-10) &         1.06(-10) &         1.02(-10) &         1.01(-10) \vspace{0.2cm} \\
$1_{ 1,  0 }  \to 0_{ 0,  0 } $  &         5.56(-11) &         6.65(-11) &         7.22(-11) &         6.93(-11) &         6.79(-11) &         6.80(-11) \\
$1_{ 1,  0 }  \to 1_{ 0,  1 } $  &         1.46(-11) &         1.38(-11) &         9.61(-12) &         6.05(-12) &         4.56(-12) &         3.89(-12) \\
$1_{ 1,  0 }  \to 1_{ 1,  1 } $  &         1.55(-11) &         1.52(-11) &         1.16(-11) &         8.35(-12) &         6.79(-12) &         5.90(-12) \vspace{0.2cm} \\
$2_{ 0,  2 }  \to 0_{ 0,  0 } $  &         9.34(-12) &         6.89(-12) &         4.68(-12) &         3.24(-12) &         2.62(-12) &         2.29(-12) \\
$2_{ 0,  2 }  \to 1_{ 0,  1 } $  &         2.76(-11) &         2.73(-11) &         2.70(-11) &         2.61(-11) &         2.58(-11) &         2.60(-11) \\
$2_{ 0,  2 }  \to 1_{ 1,  1 } $  &         1.79(-11) &         1.42(-11) &         9.32(-12) &         5.85(-12) &         4.39(-12) &         3.70(-12) \\
$2_{ 0,  2 }  \to 1_{ 1,  0 } $  &         2.28(-11) &         2.17(-11) &         1.84(-11) &         1.47(-11) &         1.30(-11) &         1.22(-11) \vspace{0.2cm} \\
$2_{ 1,  2 }  \to 0_{ 0,  0 } $  &         2.00(-12) &         1.89(-12) &         1.29(-12) &         7.45(-13) &         4.94(-13) &         3.58(-13) \\
$2_{ 1,  2 }  \to 1_{ 0,  1 } $  &         7.80(-12) &         7.14(-12) &         5.16(-12) &         3.42(-12) &         2.58(-12) &         2.14(-12) \\
$2_{ 1,  2 }  \to 1_{ 1,  1 } $  &         2.69(-11) &         2.71(-11) &         2.60(-11) &         2.48(-11) &         2.46(-11) &         2.47(-11) \\
$2_{ 1,  2 }  \to 1_{ 1,  0 } $  &         1.25(-11) &         1.05(-11) &         6.61(-12) &         3.85(-12) &         2.76(-12) &         2.27(-12) \\
$2_{ 1,  2 }  \to 2_{ 0,  2 } $  &         9.57(-11) &         9.58(-11) &         1.05(-10) &         1.09(-10) &         1.11(-10) &         1.13(-10) \vspace{0.2cm} \\
$2_{ 1,  1 }  \to 0_{ 0,  0 } $  &         5.35(-12) &         4.60(-12) &         3.43(-12) &         2.50(-12) &         2.14(-12) &         2.04(-12) \\
$2_{ 1,  1 }  \to 1_{ 0,  1 } $  &         5.38(-11) &         5.58(-11) &         6.03(-11) &         6.10(-11) &         6.13(-11) &         6.20(-11) \\
$2_{ 1,  1 }  \to 1_{ 1,  1 } $  &         2.90(-11) &         3.06(-11) &         2.92(-11) &         2.44(-11) &         2.10(-11) &         1.88(-11) \\
$2_{ 1,  1 }  \to 1_{ 1,  0 } $  &         8.99(-12) &         7.69(-12) &         5.67(-12) &         4.23(-12) &         3.57(-12) &         3.22(-12) \\
$2_{ 1,  1 }  \to 2_{ 0,  2 } $  &         1.48(-11) &         1.47(-11) &         1.16(-11) &         8.26(-12) &         6.95(-12) &         6.54(-12) \\
$2_{ 1,  1 }  \to 2_{ 1,  2 } $  &         2.35(-11) &         2.37(-11) &         2.33(-11) &         2.26(-11) &         2.25(-11) &         2.28(-11) \vspace{0.2cm} \\
$2_{ 2,  1 }  \to 0_{ 0,  0 } $  &         8.19(-13) &         6.91(-13) &         4.63(-13) &         2.76(-13) &         1.87(-13) &         1.37(-13) \\
$2_{ 2,  1 }  \to 1_{ 0,  1 } $  &         2.36(-11) &         2.51(-11) &         2.60(-11) &         2.27(-11) &         1.99(-11) &         1.81(-11) \\
$2_{ 2,  1 }  \to 1_{ 1,  1 } $  &         5.44(-11) &         5.61(-11) &         6.01(-11) &         6.14(-11) &         6.23(-11) &         6.35(-11) \\
$2_{ 2,  1 }  \to 1_{ 1,  0 } $  &         3.14(-12) &         2.78(-12) &         2.10(-12) &         1.48(-12) &         1.21(-12) &         1.10(-12) \\
$2_{ 2,  1 }  \to 2_{ 0,  2 } $  &         3.05(-11) &         2.54(-11) &         1.97(-11) &         1.66(-11) &         1.55(-11) &         1.50(-11) \\
$2_{ 2,  1 }  \to 2_{ 1,  2 } $  &         2.33(-11) &         1.91(-11) &         1.31(-11) &         9.46(-12) &         8.31(-12) &         8.10(-12) \\
$2_{ 2,  1 }  \to 2_{ 1,  1 } $  &         3.16(-11) &         3.87(-11) &         4.37(-11) &         4.23(-11) &         4.12(-11) &         4.09(-11) \vspace{0.2cm} \\
$2_{ 2,  0 }  \to 0_{ 0,  0 } $  &         1.63(-11) &         1.58(-11) &         1.43(-11) &         1.21(-11) &         1.06(-11) &         9.73(-12) \\
$2_{ 2,  0 }  \to 1_{ 0,  1 } $  &         6.02(-12) &         5.20(-12) &         4.03(-12) &         3.11(-12) &         2.68(-12) &         2.46(-12) \\
$2_{ 2,  0 }  \to 1_{ 1,  1 } $  &         3.85(-12) &         3.24(-12) &         2.46(-12) &         1.87(-12) &         1.63(-12) &         1.57(-12) \\
$2_{ 2,  0 }  \to 1_{ 1,  0 } $  &         7.38(-11) &         7.42(-11) &         7.48(-11) &         7.43(-11) &         7.48(-11) &         7.60(-11) \\
$2_{ 2,  0 }  \to 2_{ 0,  2 } $  &         2.24(-11) &         1.98(-11) &         1.57(-11) &         1.20(-11) &         1.00(-11) &         8.82(-12) \\
$2_{ 2,  0 }  \to 2_{ 1,  2 } $  &         1.57(-11) &         1.72(-11) &         1.88(-11) &         1.85(-11) &         1.82(-11) &         1.82(-11) \\
$2_{ 2,  0 }  \to 2_{ 1,  1 } $  &         1.33(-11) &         1.25(-11) &         8.88(-12) &         5.83(-12) &         4.48(-12) &         3.84(-12) \\
$2_{ 2,  0 }  \to 2_{ 2,  1 } $  &         9.50(-12) &         8.68(-12) &         9.44(-12) &         9.41(-12) &         9.05(-12) &         8.78(-12) \\
\hline
\end{tabular}
\end{center}
\label{table:rates_j2=0-0}
\end{table*}%

\subsection{Null CS cross sections}

As outlined in Section \ref{dynamique}, the transitions $J_\tau \to 0_0$ where $\tau$ 
ranges from $\tau = -J+1$ to $\tau = J-1$ with a step of 2 are predicted to be equal to zero with the CS formalism.
Using the available CC calculations, we are however able to calculate these rate coefficients for the transitions with  
$J \leq  3$ and up to 300 K with a reasonable accuracy (see Appendix B).
Since the current set of rate coefficients considers levels up to $J_\tau = 7_7$, it is necessary to estimate the rate
coefficients for the transitions that connect the fundamental energy level to the levels with $J \geq 4$ 
which are predicted to be null with the CS formalism. To do so, we extrapolate the behaviour observed for the rates
with levels such that $J < 4$ and derive the following expression :
\begin{eqnarray}
R(J_{\tau_i} \to 0_0) & = & R(1_0 \to 0_0) \times \left( \frac{1}{1.35} \right)^{J-1} \times \left( \frac{1}{2.6} \right)^{i-1} \\
\textrm{with} & & \quad \tau_i = -J+1 + 2 \times(i-1) \quad ; \quad i \in \llbracket 1 ; n \rrbracket \nonumber \\ 
\textrm{and}  & &  \quad n = E\left(\frac{2J+1}{2}\right)  \nonumber 
\end{eqnarray}
We can expect large errors of the corresponding rate coefficients, typically of a factor 
10 or higher, since this expression is established on the basis of a poor statistics. 
However, since the corresponding rate coefficients should play only a minor role in the pumping scheme, 
these uncertainties should not have noticeable consequences on the predicted line intensities. 
In any astrophysical application, this assumption should however be checked.
For example, as was done in \citet{daniel2013}, the sensitivity of the line intensities to the corresponding rate
coefficients can be assessed by considering randomly scaled values for these rates around their nominal value.

\subsection{Rate coefficients accuracy}

The accuracy of the current rate coefficients is limited by the following assumptions. 
A first source of error comes from the neglect of the inversion motion. However, 
we expect that this hypothesis should only have marginal effects on the rate coefficients, 
at least in comparison to the other sources of error.
A second limitation comes from the fact that the PES was calculated for the equilibrium
geometry of the NH$_3$ molecule. The PES was corrected for the displacement of the center of mass
between NH$_3$ and NH$_2$D, but the variation of the bond length from NH to ND was neglected.
The effect introduced by taking into account the vibrationally averaged geometry of NH$_2$D, rather than taking the 
NH$_3$ internal geometry, can lead to noticeable effects. Such a modification was tested for 
D$_2$O by \cite{scribano2010}, where rate coefficients for this isotopologue were calculated
using the H$_2$O internal geometry. Differences of the order of 30\% were found and we might 
expect differences of the same order for NH$_2$D. 
Finally, a third source of error resides in the use of the CS formalism at high temperatures.
As described in Sec. \ref{dynamique} and in Appendix A, the CS calculations were scaled according to some reference
calculations performed with the CC formalism. However, an inherent limitation of this procedure comes 
from the uncertainty introduced by the scattering of the ratios which are due to the resonances. Moreover,
the ratios were extrapolated at the highest energies. Finally, for the transitions that involve energy levels 
higher than the $4_2$ level, we directly computed the rate coefficients using the original CS cross sections.

\begin{figure}
\begin{center}
\includegraphics[angle=270,scale=0.43]{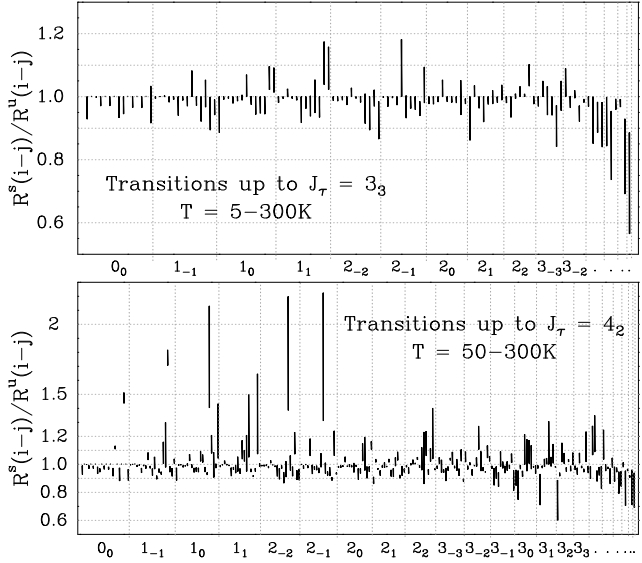}
\caption{\textbf{Upper panel}: range of the variations found in the temperature range $T = 5-300$ K, 
between the scaled and unscaled collisional rate coefficients for the levels up to $J_\tau = 3_3$.
The x--axis gives the final level of the de--excitation collisional rate coefficient. 
For each final level, the variations are given for all the de--excitation rates, ordered 
by increasing value of the upper energy level state. \textbf{Lower panel}: same as the upper panel 
but for the transitions which consider levels up to $J_\tau = 4_2$ and for the temperature range $T = 50-300$ K.}
\label{fig:accuracy}
\end{center}
\end{figure}

In order to have an estimate of the error introduced in the rate coefficients due to the use
of the CS formalism, we compared the rate coefficients calculated with the scaled CS cross sections 
(noted $R^s(i \to j)$) to the rates calculated with the unscaled cross sections (noted $R^u(i \to j)$). The minimum
and maximum values of the ratios $R^s(i \to j)/R^u(i \to j)$ found in the range $T = 5-300$ K are reported 
in the upper panel of Fig. \ref{fig:accuracy} for all the de-excitation transitions up to $J_\tau = 3_3$. 
In the lower panel of this figure, we report the same quantities, but for levels up to $J_\tau = 4_2$
and for the temperature range $T = 50-300$ K.
From this figure, it can be seen that for the transitions that involve levels up to the $3_{-1}$ (E$\sim$75 cm$^{-1}$), 
the scaled and unscaled rate coefficients differ by less than 20\% (upper panel). 
On the other hand, for higher energy levels, the differences can be as high as a factor $\sim$2 but
most of the transitions (93\% of the 253 transitions considered here) show differences below a factor 1.3.
We note that the reason why the lowest energy levels are less affected by the scaling of the cross 
sections comes from the fact that a larger part of the grid correspond to CC calculations for these levels. 
Finally, given the variations observed for the $R^s(i \to j)/R^u(i \to j)$ ratios, we can expect that the transitions
that involve an energy level above $J_\tau = 4_2$ will have a typical accuracy of 30\%, since these transitions
are directly calculated from the CS cross sections. \\

To summarize the previous considerations, considering the current PES,
we expect that the current rate coefficients will be accurate within 5\% for the transitions
up to $J_\tau = 4_2$. For higher energy levels, the accuracy should be of the order of 30\%. 
This holds true for all transitions except for the transitions  $J_\tau \to 0_0$ where $\tau$ 
ranges from $\tau = -J+1$ to $\tau = J-1$ with a step of 2. For these transitions and for the levels
such that $J<4$, the rates should be accurate to within 30\%. For higher levels, i.e. with $J \geq 4$, 
we can expect an error up to a factor 10 in the rate coefficients. Most importantly, a dominant
source of error resides in the fact that the PES was calculated for a molecular geometry
which corresponds to the NH$_3$ internal geometry. Thus, we might expect an additional 
30\% uncertainty on the rate coefficients \citep{scribano2010}.

\subsection{Thermalized rate coefficients}

\begin{figure}
\begin{center}
\includegraphics[angle=270,scale=0.6,angle=90]{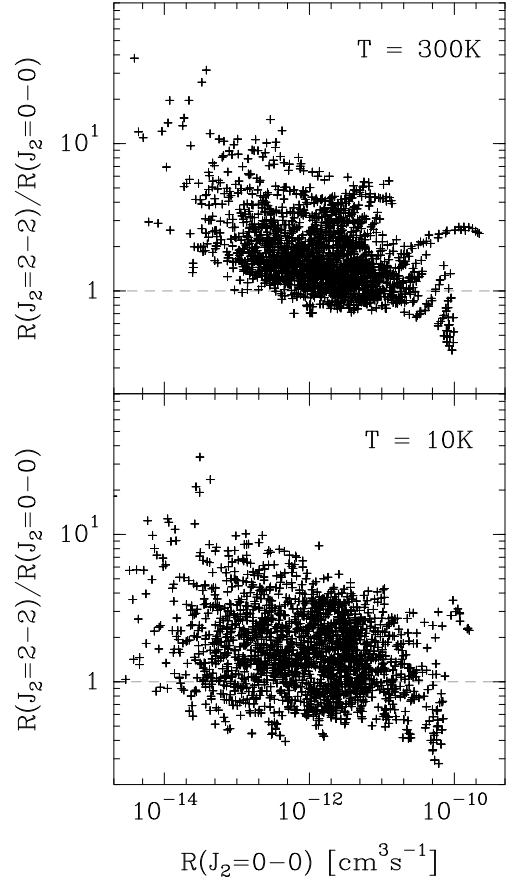}
\caption{Ratio of the rate coefficients associated to the $J_2 = 2 \to 2$ over the rate with $J_2 = 0 \to 0$,
as a function of the magnitude of the rate coefficients associated to the para--H$_2$ fundamental state. The ratios
are given at T= 10 K (lower panel) and T = 300K (upper panel).}
\label{fig:ratio_J2=2}
\end{center}
\end{figure}

The previous discussion dealt with rate coefficients where para--H$_2$ was restricted 
to its fundamental state $J_2 = 0$. Such rate coefficients are enough to tackle astrophysical
applications that deal with cold media (i.e. $T < 50K$). However, at higher gas 
temperatures, we expect a larger fraction of the H$_2$ molecules in excited states ($J_2 > 0$). 
As discussed in Sec. \ref{dynamique}, the cross sections with $J_2 = 2 \to 2$ are of larger 
magnitude than the cross sections associated with the $J_2 = 0 \to 0$ transitions. 
This is shown in Fig. \ref{fig:ratio_J2=2} where we represent the ratio of the rate coefficients with 
$J_2 = 2 \to 2$ to that with $J_2 = 0 \to 0$, for temperatures of 10 and 300K. On average,
the rate coefficients with $J_2 = 2 \to 2$ are of larger magnitude, with an average ratio slightly increasing
from 2 to 2.4 with temperature increasing from 10K to 300K. 
However, as can be seen from Fig. \ref{fig:ratio_J2=2}, the variation in magnitude of the rate 
coefficient depends on the absolute magnitude of the rate, the variation being 
larger for the rates of lower magnitude. Moreover, as seen in Fig. \ref{fig:ratio_J2=2}, 
there is a large scatter of the ratios.
Thus, by varying
the relative populations of the fundamental and first H$_2$ excited states, the number of 
collisionnally induced events will be modified accordingly, i.e. the number of events 
becomes larger with increasing the fractional abundance of the first excited state. 
To take this effect into account, we define thermalized rate coefficients, following the 
definition given in \citet{dubernet2009} (i.e. Eq. (4) in this reference).

Finally, we stress that in hot media, the relevant collisional partner will 
be ortho--H$_2$ rather than para--H$_2$ and in such a case, it should be necessary to calculate
the corresponding rate coefficients. However, in the case of the 
H$_2$O / H$_2$ collisional system it was found that the effective rate coefficients for $J_2 = 2$ 
and $J_2 = 1$ agree within $\sim$30\% \citep{dubernet2009,daniel2010,daniel2011}. 
More generally, these studies showed that apart from the rate coefficients in $J_2 = 0$, 
all the effective rate coefficients with $J_2 > 0$ are qualitatively similar, independently 
of the H$_2$ symmetry. As a consequence, in astrophysical applications,
the thermalized rate coefficients with ortho--H$_2$ or para--H$_2$ lead to similar 
line intensities at high temperatures \citep{daniel2012}.
To conclude, this means that in principle, the rate coefficients with $J_2 = 2 \to 2$ can serve as a template 
to emulate the rate coefficients of ortho--H$_2$ ($J_2 = 1 \to 1$). 
We checked this assumption by performing a few calculations with ortho--H$_2$ as a collisional
partner. In Fig. \ref{fig:XS_J2=2}, we compare the cross sections of the $2_{-2} - 1_0$ transition associated with either
$J_2 = 2 \to 2$ or $J_2 = 1 \to 1$. It can be seen that these two cross sections are qualitatively similar for the energy
range considered here. Such an agreement is also found on a more general basis.  
Considering a statistics over the $i \to j$ transitions, 
we obtained that, whatever the energy, the mean value $\mu$ of the ratios $\sigma_{ij}(2 \to 2) / \sigma_{ij}(1 \to 1)$
is $\mu \sim 1$, with a standard deviation $\sigma$ that vary with energy in the range $\sigma = 0.15-0.30$.
 For each energy, $\sim$70\% of the transitions vary by less 
than $1\, \sigma$ around $\mu$ and $\sim$95\% of the transitions are within $2 \, \sigma$.
Thus, this confirms that the rate coefficients in  $J_2 = 2 \to 2$ are a good approximation for the ortho--H$_2$ rate coefficients.

\section{Discussion} \label{discussion}

\begin{figure}
\begin{center}
\includegraphics[angle=0,scale=0.43]{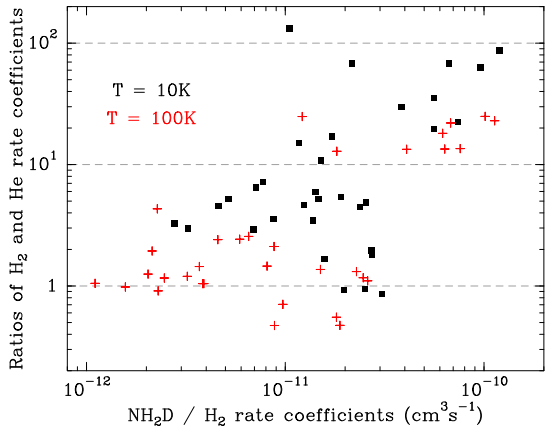}
\caption{Comparison between the current NH$_2$D / H$_2$ rate coefficients with 
the rates calculated with He as a collisional partner \citep{machin2006}.}
\label{fig:comp-He}
\end{center}
\end{figure}

Prior to this work, the only available rate coefficients for NH$_2$D were 
calculated with He as a collisional partner. Rate coefficients with He 
are often used as a template to estimate the rate coefficients with H$_2$. 
However, for many molecules, large differences exist and the rates 
do not scale accordingly to the factor $\sim$1.4 deduced from the difference in 
the reduced masses of the two colliding systems. Concerning the deuterated
isotopologues of ammonia, such a conclusion was already reached for ND$_2$H.
Indeed, for this particular isotopologue, the He and H$_2$ rate coefficients 
were found to vary by factors in the range 3--30 \citep{machin2007,wiesenfeld2011}.
In Fig. \ref{fig:comp-He}, we give the ratios of the current NH$_2$D / H$_2$ rate coefficients
with the rates calculated with He \citep{machin2006}, as a function of the magnitude of the 
H$_2$ rate coefficients and for temperatures of T = 10K and T = 100K. 
The comparison concerns the transitions that involve levels up to $J_\tau = 2_2$  \cite[the highest
level considered in][]{machin2006}.
From this figure, it appears that the H$_2$ and He rate coefficients 
show large variations, as expected from the previous studies of ND$_2$H.
At T = 10K, the ratios are in the range 1--100. Moreover, it appears that the largest
differences are also found for the largest rate coefficients. At higher temperatures, 
we obtain the same trend except that the range of variations is reduced. For example,
at T = 100K, the ratios for the lowest rate coefficients are around 1 while they are around 
20 for the highest rates.

\begin{figure}
\begin{center}
\includegraphics[angle=0,scale=0.5]{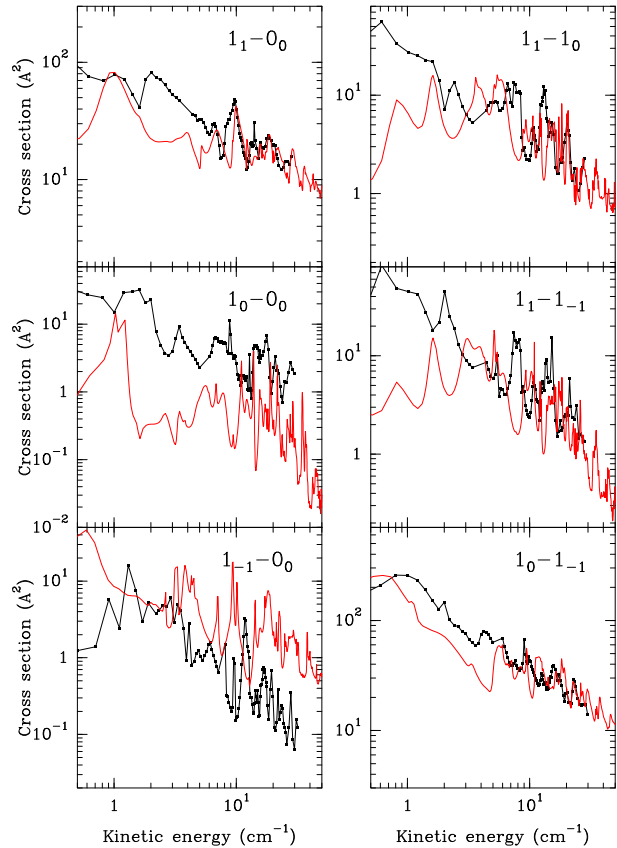}
\caption{Comparison between the NH$_2$D (red) and ND$_2$H (black) cross sections calculated
with H$_2$ ($J_2 = 0,2$).}
\label{fig:comp_NH2D-ND2H}
\end{center}
\end{figure}

Finally, one may wonder if it is possible to infer the rate coefficients of singly
deuterated ammonia from the rates of doubly deuterated ammonia, or vice--versa. 
A direct comparison of the current rates with the values reported by 
\citet{wiesenfeld2011} for ND$_2$H / H$_2$ would not be totally meaningful, at least
quantitatively, since in the latter study, the dynamical calculations were performed 
with a $J_2$ = 0 basis for H$_2$. As discussed in Sec. \ref{dynamique}, 
considering a $J_2$ = 0,2 basis for H$_2$ can induce variations as large as a 
factor $\sim$7 and the magnitude of the effects are presumably similar for ND$_2$H.
We thus performed a few additional calculations for the ND$_2$H / H$_2$ system. 
These calculations are similar to the ones presented in \citet{wiesenfeld2011}
except for the H$_2$ basis which is now set to $J_2 = 0, 2$. 
In Fig. \ref{fig:comp_NH2D-ND2H}, we compare the cross sections of NH$_2$D 
and ND$_2$H for the six de--excitation transitions possible when considering 
the levels up to $J_\tau = 1_1$. From this figure, it appears that on the average, 
the cross sections are of the same order of magnitude for the two colliding systems.
The differences are definitely lower than those found with He as 
a collisional partner. However, above 5 cm$^{-1}$, differences of a factor $\sim$3 or higher 
may occur when considering individual transitions. This implies that a dedicated calculation
is necessary to infer accurate rate coefficients for a particular isotopologue. 
Furthermore, given the differences obtained between the rate coefficients calculated with 
H$_2$ or He, the current comparison shows that in the absence of a 
dedicated calculation, the rate coefficients of a deuterated isotopologue
should be preferentially taken from another isotopologue with the same symmetry rather than scaled from 
the He rates. As an example, to emulate the CD$_2$H / H$_2$ rate coefficients, it should
be more accurate to take the  CH$_2$D / H$_2$ rate coefficients than the 
CD$_2$H / He one.

\section{Conclusions} \label{conclusion}

We reported the calculation of collisional rate coefficients for the NH$_2$D / H$_2$ system
for NH$_2$D rotational energy levels up to $J_\tau = 7_7$ ($E \sim 735$K) and for temperatures in the range
T = 5--300K. These rate coefficients will be made available through the 
BASECOL \citep{dubernet2013} and LAMDA \citep{schoier2005} databases.
The current rate coefficients were calculated by adapting the 
NH$_3$ / H$_2$ potential energy surface presented in \citet{maret2009}. The dynamical calculations
were performed at the close-coupling level at low total energy ($E < 110$ cm$^{-1}$) and with the 
coupled--states formalism at higher energies. For the levels up to $J_\tau = 4_2$ the CS cross sections
were scaled according to some reference points performed with the accurate CC formalism. For these 
transitions and for the current PES,
the accuracy of the rate coefficients is expected to be of the order of 5\%. For higher energy 
levels, the accuracy is limited by the use of the CS formalism, and the corresponding rate coefficients
are expected to be accurate within 30\%. These uncertainties concern all the transitions except the 
$J_\tau \to 0_0$ transitions with $J_\tau$ varying from $-J+1$ to $J-1$ in a step of two. For these latter
transitions, we expect the rate coefficients to be accurate within 30\% for the levels with $J\leq3$
while for the levels such that $J \geq 4$, the rate coefficients have uncertainties of the order of a factor 10.
We note that these errors only consider the accuracy of the dynamical calculations and do no take into 
account other sources of uncertainty like the accuracy of the PES or the neglecting of the inversion motion.
For what concerns the PES, we might expect an additional 30\% uncertainty on the rate coefficients,
which is to be added in quadrature to the previously mentioned uncertainties,
since the current PES was calculated for an internal geometry suitable to NH$_3$. 

\section*{Acknowledgments}
All (or most of) the computations presented in this paper were performed using the CIMENT infrastructure (https://ciment.ujf-grenoble.fr), which is supported by the Rh\^one-Alpes region (GRANT CPER07\_13 CIRA: http://www.ci-ra.org).
D.L. support for this work was provided by NASA (\emph{Herschel} OT funding) through an award issued by JPL/Caltech.
This work has been supported by the Agence Nationale de la Recherche
(ANR-HYDRIDES), contract ANR-12-BS05-0011-01 and by the CNRS national program
``Physico-Chimie du Milieu Interstellaire''.

\appendix

\section[]{Scaling of the CS cross--sections}

\begin{figure*}
\begin{center}
\includegraphics[angle=270,scale=0.75]{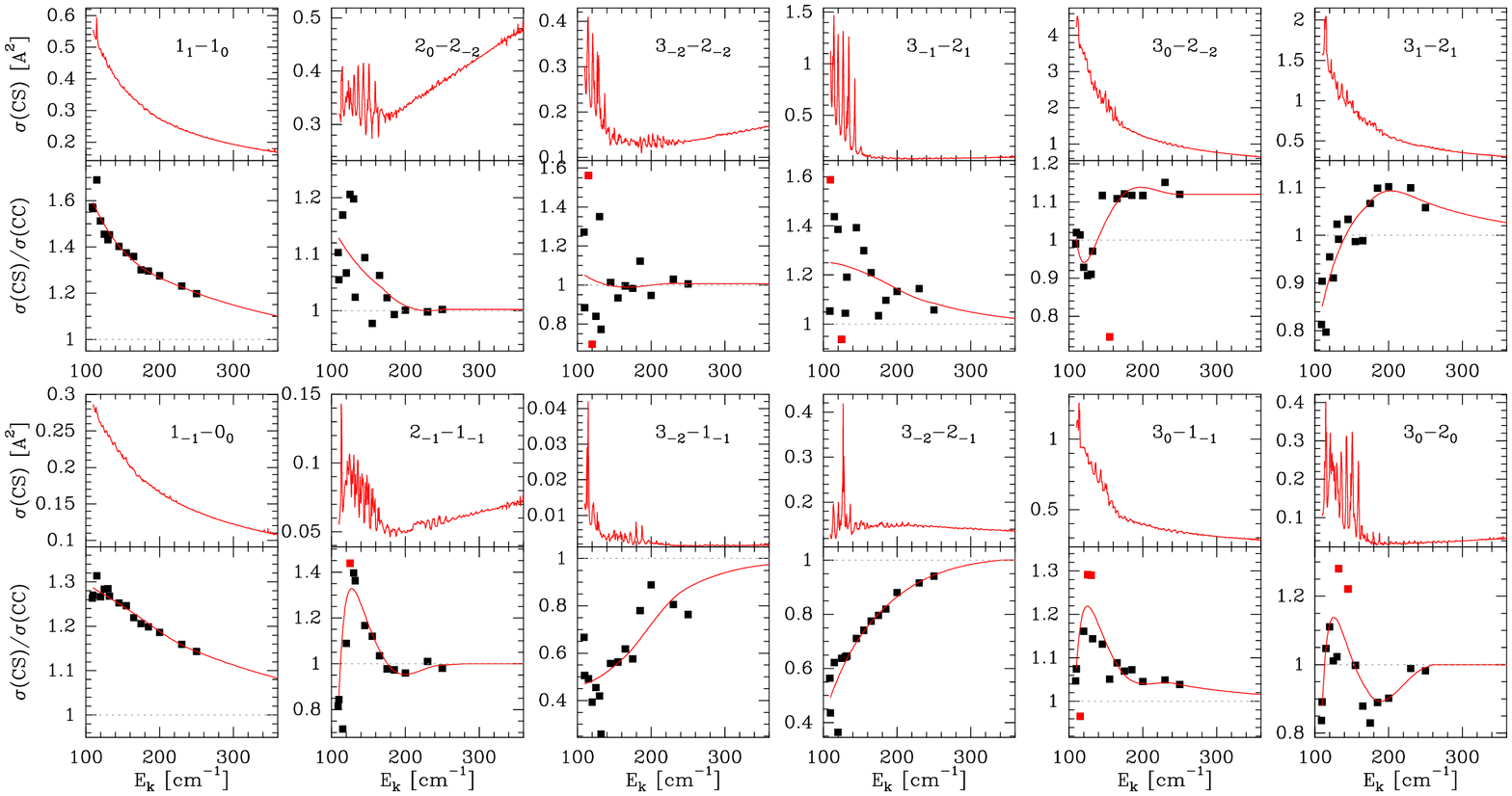}
\caption{In each panel, the upper plot shows the unscaled CS cross sections in $\AA^2$. In the
lower plot, the black dots correspond to the ratios of the CC and CS cross sections and the red curves show a fit of these ratios. }
\label{fig:scaling}
\end{center}
\end{figure*}

In order to obtain accurate cross sections, 
we scaled the CS cross sections according to a few calculations performed with the CC formalism.
These CC calculations were performed for total energies in the range 110 cm$^{-1}$ $<$ E $<$ 250 cm$^{-1}$.
The ratio of the CC and CS cross sections is reported in Fig. \ref{fig:scaling} for a few transitions.
In order to correct the CS cross sections, we performed an analytical fit of the ratio and to that purpose,
we used two fitting functions : 
\begin{eqnarray}
f(E) & = & \sum_{i\in\{1,3\}} a_i \, E^{i-2} \\
g(E) & = & \sum_{i\in\{1,3\}} a_i \, E^{i-2} + a_4 \, e^{-0.1 \, E} 
\end{eqnarray}
where $E$ corresponds to the total energy. The latter functional form was used to reproduce the 
oscillating behaviour seen in some transitions (like e.g. the transitions $2_{-1}$--$1_{-1}$, $3_{0}$--$1_{-1}$ or 
$3_{0}$--$2_{0}$ shown in Fig. \ref{fig:scaling}). The choice of the functional form is made automatically, 
and for a given transition the function $g(E)$ is selected as soon as the $\chi^2$ obtained with this function 
is lower by at least a factor 3 by comparison to the $\chi^2$ obtained with the function $f(E)$.

\begin{figure}
\begin{center}
\includegraphics[angle=0,scale=0.5]{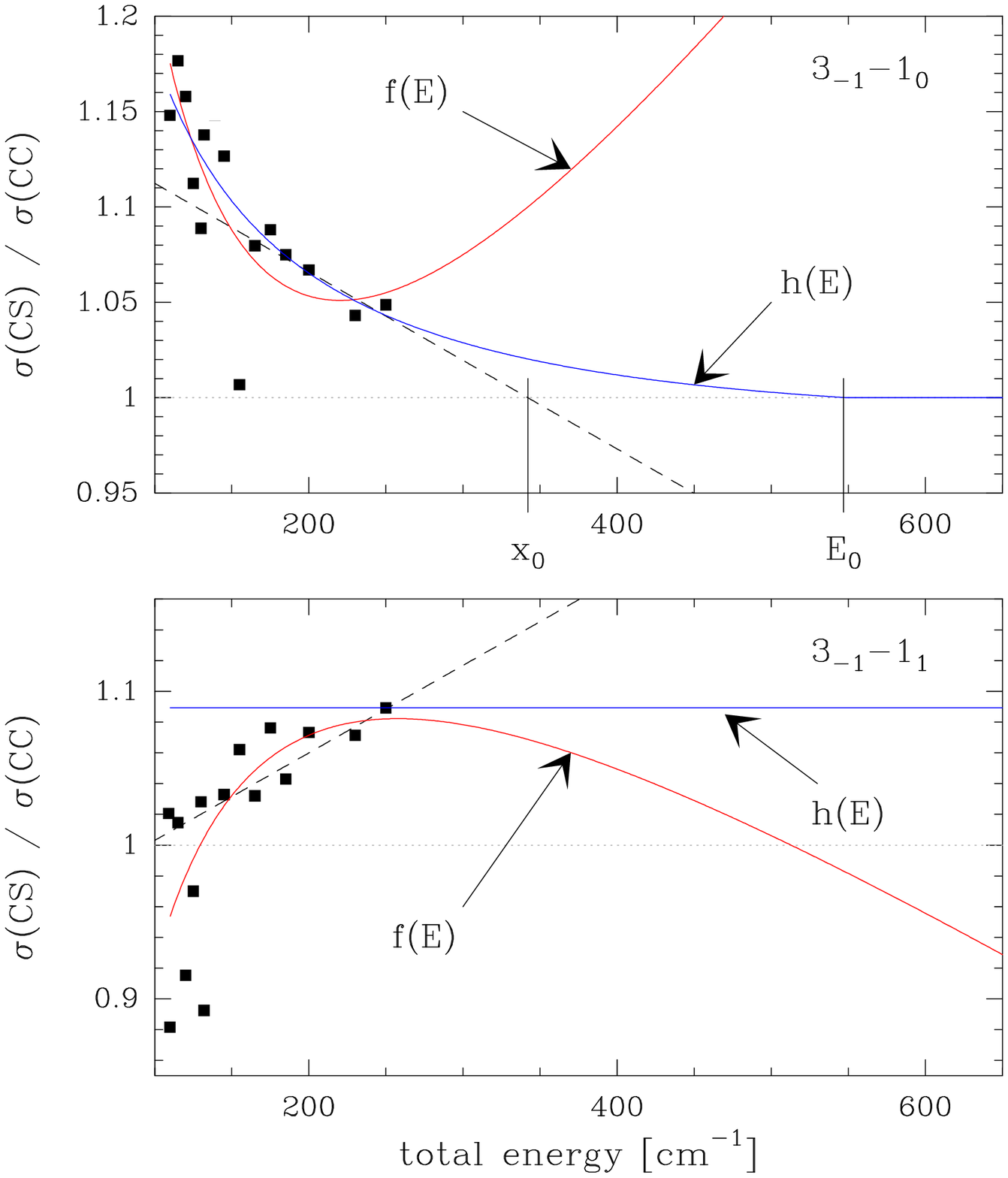}
\caption{Extrapolation of the CS / CC ratios (see text for details).}
\label{fig:extrapol}
\end{center}
\end{figure}

Considering the behaviour of the ratios reported in Fig. \ref{fig:scaling}, it appears that the ratios depend 
on the total energy and, in most cases, tend to get close to 1 at the highest energies. Since we are interested in scaling the cross
sections up to $\sim$ 3000 cm$^{-1}$, we extrapolate the ratios at energies higher than 250 cm$^{-1}$. 
To do so, we proceed as follow. We first fit a linear function using the ratios obtained at the highest energies. This 
leads to two possible behaviours depending on the energy where the resulting line crosses the $y = 1$ axis
(denoted $x_0$ in what follows).
In the first case, x$_0$ $>$ 250 cm$^{-1}$, and we then use the $f(E)$ functional 
form to extrapolate the ratios at high energy. In the fitting procedure, we put as an additional constraint 
that the ratio is 1 at $E_0 = 1.5 \, x_0$. Above $E_0$, the ratio is set to one.
In the second case, x$_0$ $<$ 250 cm$^{-1}$. In this case, we fix the ratio above 250 cm$^{-1}$
and keep the value obtained at this latest energy. 
This extrapolation procedure thus gives the function $h(E)$.
These two cases are illustrated in Fig. \ref{fig:extrapol} where
we show the result of the fitting procedure for the $3_{-1}-1_{0}$ transition (case 1) 
and $3_{-1}-1_{-1}$ transition (case 2). 

Finally, to insure the continuity between the two fitting functions, we define the function $\phi(E)$, 
which gives the correction of the CS cross sections in the range 110 $<$ E $<$ 3000 cm$^{-1}$: 
\begin{eqnarray}
\phi(E) & = & f(E) \, \textrm{ or } \, g(E) \quad \textrm{if} \quad E < 170 \, \textrm{cm}^{-1} \\
\phi(E) & = & (1-w) \, f(E) +  w \, h(E) \quad \textrm{if} \quad 170 < E < 250 \, \textrm{cm}^{-1} \\
\phi(E) & = & h(E) \quad \textrm{if} \quad E > 250 \, \textrm{cm}^{-1}
\end{eqnarray}
where $w$ is a weighting function that vary linearly between 0 and 1 in the range 170 $<$ E $<$ 250 cm$^{-1}$.
The resulting fit is shown in Fig. \ref{fig:scaling} for a few transitions which are characteristic of the behaviour 
observed on the whole sample of transitions. 


\section{Null CS cross sections}

\begin{figure}
\begin{center}
\includegraphics[angle=270,scale=0.55]{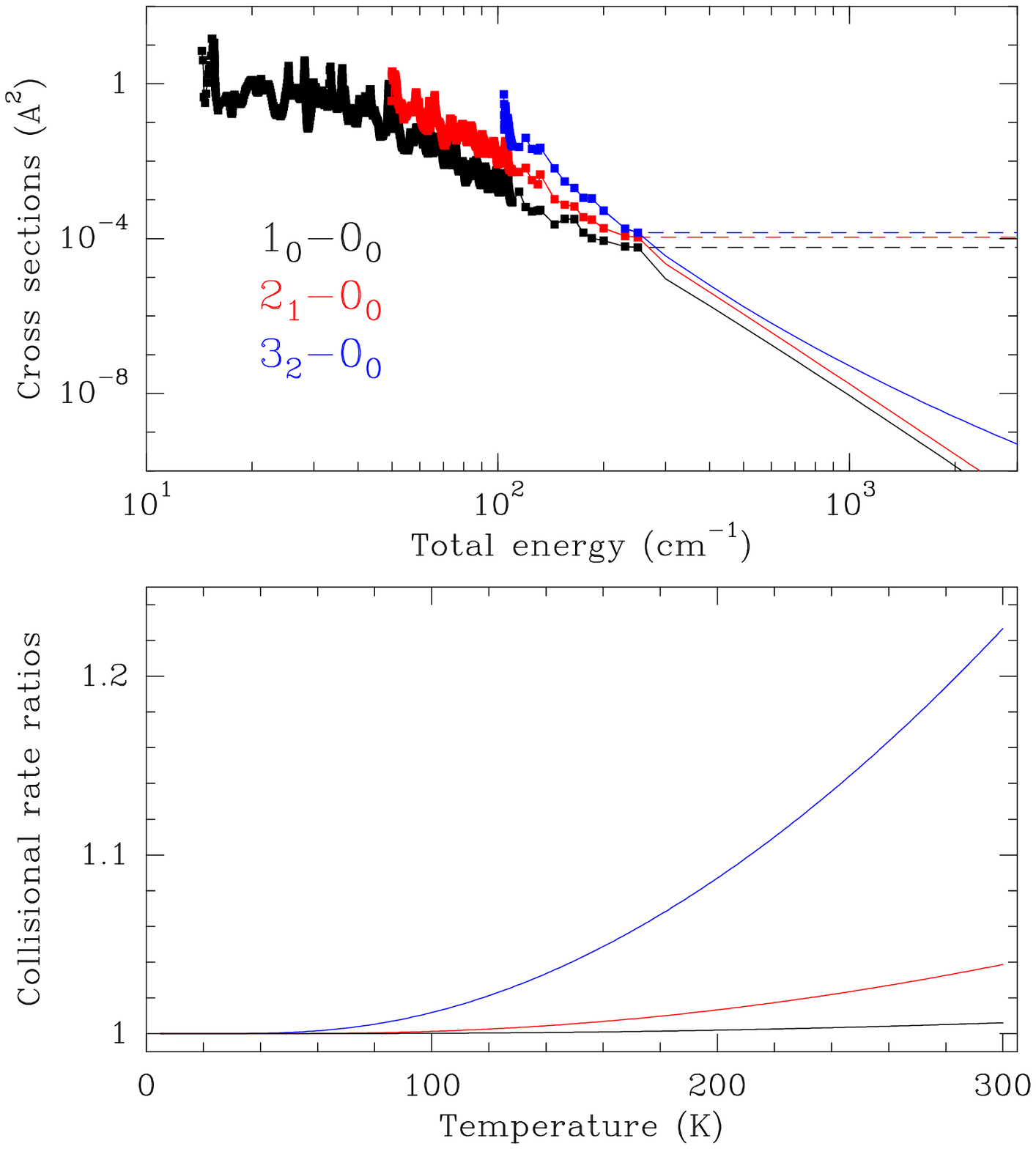}
\caption{The upper panel gives the CC cross sections of the $1_0 \to 0_0$, $2_1 \to 0_0$ and $3_2 \to 0_0$
transitions as a function of the total energy. The plain curves correspond to an extrapolation of these cross section
based on a polynomial fit. The dashed curves, which assume constant values above 250 cm$^{-1}$, 
would correspond to an upper limit of the cross sections at high energies. The lower panel give the ratios of the rate coefficients obtained with the two extrapolations considered in the upper panel (same color code).}
\label{fig:ratios_rates}
\end{center}
\end{figure}

As outlined in Section \ref{dynamique} and \ref{rates}, the transitions $J_\tau \to 0_0$ where $\tau$ 
ranges from $\tau = -J+1$ to $\tau = J-1$ in a step of 2 are predicted to be equal to zero with the CS formalism.
Since we performed CC calculations below 110 cm$^{-1}$ with some additional points between
110 and 250 cm$^{-1}$, we are however able to calculate the rate coefficients for a few of these transitions
and up to 300 K, despite the lack of calculated points above 250 cm$^{-1}$. This is due to the fact that the 
corresponding cross sections decrease sharply with the kinetic energy. This is illustrated in Fig. \ref{fig:ratios_rates}
for the $1_0 \to 0_0$, $2_1 \to 0_0$ and $3_2 \to 0_0$ transitions.
The upper panel in this figure shows the available CC points as well as two possible extrapolations of the cross sections
above 250 cm$^{-1}$. The first extrapolation is obtained using a polynomial fit of the CC cross sections (plain curves)
while the second way of extrapolating takes a constant value for the cross sections above a total energy of 250 cm$^{-1}$
(dashed curves). These two limiting cases seem reasonable in view of the overall behaviour of the cross sections 
shown in Fig. \ref{fig:CS_null}. Using these two extrapolations, we then calculated the corresponding 
rate coefficients for $T = 5-300$ K and their ratios are indicated in the bottom panel of Fig. \ref{fig:ratios_rates}.
It can be seen that for these transitions, the rate coefficients are identical, within 2\%, below $T = 100$ K. For larger
temperatures, we obtain larger differences, the discrepancies being the highest at 300 K. 
However, the variations are always below 25\% and this conclusion holds true for the 7 transitions that consider 
the levels below the $J_\tau = 3_{2}$ level.

\begin{figure}
\begin{center}
\includegraphics[angle=0,scale=0.48]{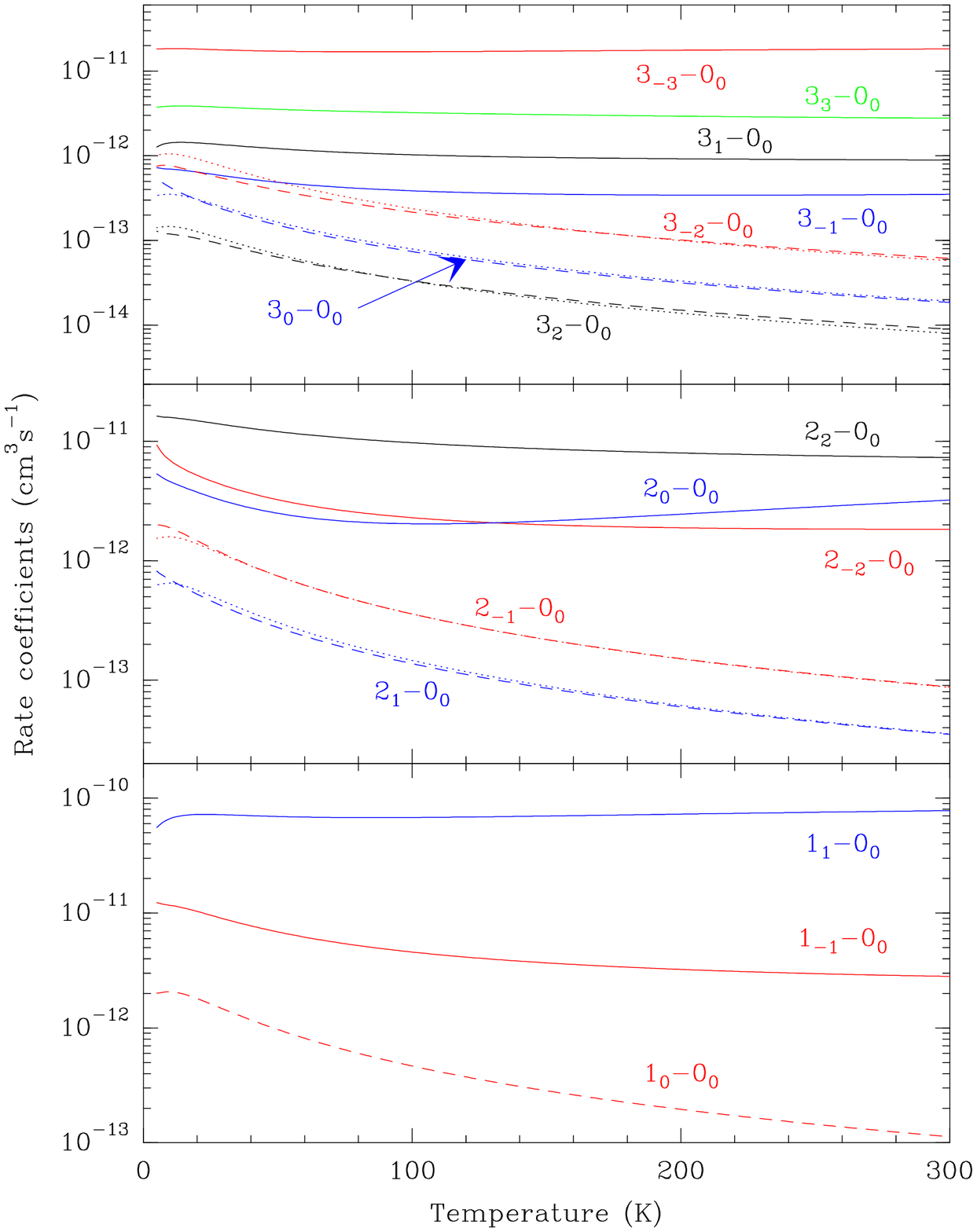}
\caption{Rate coefficients for the transitions connecting the levels from $J = 1$ (lower panel),
$J = 2$ (middle panel) and $J = 3$ (upper panel) to the fundamental energy level. The transitions
for which the CS formalism predict non null rate coefficients are indicated in plain curves. For the other
transition for which the CS formalism predict null cross sections, the dashed curves correspond to the rate
coefficients calculated from the available CC points. In the latter case, for the transitions from 
$J = 2$ and $J = 3$, we also show rate coefficients which are scaled from the 
$1_0 \to 0_0$ transition (dotted curves).}
\label{fig:extrapol_CS-null}
\end{center}
\end{figure}

Since the current set of rate coefficients considers levels up to $J = 7$, it is necessary to complete the current set.
To do so, we extrapolate the behaviour observed for the rates
with levels such that $J < 4$. In Fig. \ref{fig:extrapol_CS-null}, we show the various rate coefficients calculated for the 
levels up to $J_\tau = 3_3$. First, as can be seen in
Fig. \ref{fig:CS_null}, the rate coefficients associated with the transitions predicted to be null with the CS formalism are 
systematically of lower magnitude than the rates of the other transitions. 
Thus, they should presumably play a minor role in the radiative transfer calculations.
Moreover, all these transitions show a similar trend with temperature and are roughly proportional.
As a consequence, it is possible to mimic the rate of a given transition using the rate coefficient, for example, of the 
$1_0 \to 0_0$ transition. The dotted curves in Fig. \ref{fig:extrapol_CS-null}, for the transitions from  
J = 2 and J = 3, are approximated in this way by using the analytical expressions:
\begin{eqnarray}
R(2_{-1} \to 0_0) & = & R(1_0 \to 0_0) \times \left( \frac{1}{1.3} \right)  \\
R(2_{1} \to 0_0) & = & R(1_0 \to 0_0) \times \left( \frac{1}{1.3} \right) \times  \frac{1}{2.45}  \\
R(3_{-2} \to 0_0) & = & R(1_0 \to 0_0) \times \left( \frac{1}{1.4} \right)^2  \\
R(3_{0} \to 0_0) & = & R(1_0 \to 0_0) \times \left( \frac{1}{1.4} \right)^2 \times  \frac{1}{3}  \\
R(3_{2} \to 0_0) & = & R(1_0 \to 0_0) \times \left( \frac{1}{1.4} \right)^2 \times  \frac{1}{3} \times \frac{1}{2.4}  
\end{eqnarray}
It can be seen in Fig. \ref{fig:extrapol_CS-null} that this simple scaling of the $1_0 \to 0_0$ rate coefficient
 can reproduce qualitatively 
the rate coefficients directly obtained from the integration of the cross sections (dashed curves). 
From this trend, we can generalize the above equations and the resulting expression 
corresponds to Eq. (1) of Sections \ref{rates}.

\bsp

\label{lastpage}

\end{document}